\documentclass[sigconf]{acmart}

\AtBeginDocument{%
  }

\setcopyright{rightsretained}
\copyrightyear{2024}
\acmYear{2024}
\acmDOI{10.1145/3613905.3648111}

\acmConference[CHI EA '24]{Extended Abstracts of the CHI Conference on Human Factors in
Computing Systems}{May 11--16, 2024}{Honolulu, HI, USA}

\acmBooktitle{Extended Abstracts of the CHI Conference on Human Factors in Computing
Systems (CHI EA '24), May 11--16, 2024, Honolulu, HI, USA}
\acmISBN{979-8-4007-0331-7/24/05}

\begin{document}

\title{GeoBotsVR: A Robotics Learning Game for Beginners with Hands-on Learning Simulation}

\author{Syed Tanzim Mubarrat}
\email{smubarra@purdue.edu}
\orcid{0000-0002-9702-8901}
\affiliation{%
  \institution{Purdue University}
  \streetaddress{610 Purdue Mall}
  \city{West Lafayette}
  \state{Indiana}
  \country{USA}
  \postcode{47907}
}

\renewcommand{\shortauthors}{Mubarrat}

\begin{abstract}
This article introduces \textit{GeoBotsVR}, an easily accessible virtual reality game that combines elements of puzzle-solving with robotics learning and aims to cultivate interest and motivation in robotics, programming, and electronics among individuals with limited experience in these domains. The game allows players to build and customize a two-wheeled mobile robot using various robotic components and use their robot to solve various procedurally-generated puzzles in a diverse range of environments. An innovative aspect is the inclusion of a repair feature, requiring players to address randomly generated electronics and programming issues with their robot through hands-on manipulation. \textit{GeoBotsVR} is designed to be immersive, replayable, and practical application-based, offering an enjoyable and accessible tool for beginners to acquaint themselves with robotics. The game simulates a hands-on learning experience and does not require prior technical knowledge, making it a potentially valuable resource for beginners to get an engaging introduction to the field of robotics.
\end{abstract}


\begin{teaserfigure}
  \includegraphics[width=\textwidth]{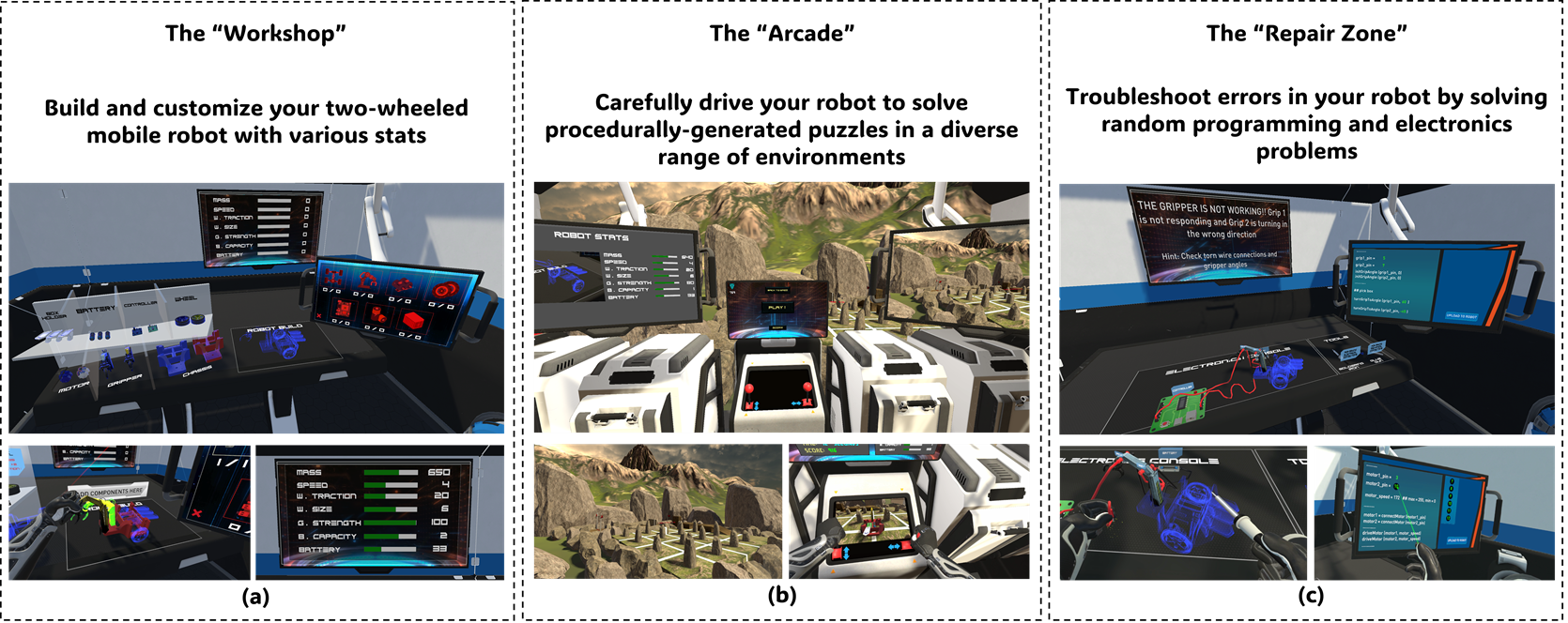}
  \caption{The three main components of GeoBotsVR: (a) the "Workshop" for building and customizing a two-wheeled mobile robot, (b) the "Arcade" for navigating a robot around a complex environment to solve puzzles, and (c) the "Repair Zone" for removing robot errors by solving electronics and programming problems}
  \Description{Screenshots showing the three main components of GeoBotsVR: the workshop, the arcade, and the repair zone.}
  \label{fig:teaser}
\end{teaserfigure}

\received{18 January 2024}
\received[accepted]{8 February 2024}

\maketitle

\section{Introduction}
In recent years, the growing emphasis on STEM (Science, Technology, Engineering, Mathematics) education has sparked increased interest in robotics \cite{rockland_advancing_2010,zeidler2016stem} as a solution to real-world problems. Knowledge and skills in the field of robotics can allow individuals to solve complex problems in various applications and apply their knowledge in real-world scenarios, spanning various timeframes, locations, and situations \cite{eguchi2014robotics}. A critical element of robot systems is the embedded system, comprised of various electronic components programmed for specific tasks \cite{watanabe2018compulsory}. However, learning embedded systems, and by extension, robotics, can be challenging due to the requisite diverse skills in programming and electronics \cite{watanabe2018compulsory}. Consequently, fostering an individual’s interest in learning robotics is difficult to attain. In this case, motivation can be a key element as it plays a crucial role in enhancing cognitive strategies \cite{ryan2000self,linnenbrink2002motivation,lynch2006motivational,mouton1984synergogy}. In addition, gamification has emerged as a promising approach to engage students in technology-related fields \cite{costa2023using,shafie2020gamification}.

Research indicates that incorporating game concepts into learning activities significantly improves students' understanding and motivation in programming \cite{costa2023using,de2018playing,lee2014principles,lee2015comparing,smiderle2020impact, kao2020hack} and electronics \cite{dewantara2021game,duran2018masterengineer,luthon2014laborem,salas2022construction}. While some studies focus on these areas separately, the integration of robotics, electronics, embedded systems, and programming within a gamified framework remains understudied. The few existing studies \cite{watanabe2018compulsory, panskyi2021holistic} often target students with prior programming and electronics education, utilizing hands-on training rather than accessible digital platforms. These approaches prioritize teaching over the enjoyment and gaming components, making them challenging for individuals inexperienced in these fields.

In terms of mode of learning, researchers suggest that virtual reality (VR) enhances learning by increasing engagement and immersion \cite{hussein2015benefits,dede2009immersive, mubarrat2020evaluation, srinivasan2021biomechanical,Chowdhury_2022_coreg,Chowdhury_2022_physics}. VR's spatial navigation reduces the cognitive load in programming learning, surpassing traditional text-based methods \cite{lee2014learning}. Additionally, VR provides a sense of self-presence \cite{ratan2013self}, fostering an embodied-cognitive learning experience \cite{melcer2018learning,shin2017role} that encourages more intuitive interaction with the content \cite{steed2016impact}, potentially improving overall learning outcomes \cite{cheon2012effects}. Therefore, I propose \textit{GeoBotsVR}\footnote{Preview video: {\href{https://youtu.be/Q0t3EfVHfrc}{\texttt{https://youtu.be/Q0t3EfVHfrc}}}}\footnote{Demo video: {\href{https://youtu.be/CBdWhMo7stQ}{\texttt{https://youtu.be/CBdWhMo7stQ}}}}, an easily accessible and easy-to-play VR puzzle-solving/racing game designed to foster interest and motivation in robotics, electronics, and embedded systems of players through an enjoyable experience. In \textit{GeoBotsVR}, the main gameplay focuses on players solving procedurally-generated puzzles using a customizable robot in different levels, while learning about robotics components, electronics, and programming along the way.

\section{Related Work}

\subsection{Gamified Approaches Towards Programming Learning}

Numerous studies have employed game-based approaches to enhance students' motivation and understanding in programming learning \cite{costa2023using,de2018playing,lee2014principles,lee2015comparing,smiderle2020impact, kao2020hack}. For instance, some studies have used a debugging game called Gidget to improve students’ computational thinking and coding proficiency \cite{lee2014principles,lee2015comparing}. Another study compared the performance of undergraduate students in a programming learning environment versus their performance in the same environment gamified through the introduction of ranking, points, and badges \cite{smiderle2020impact}. The authors found that the students in the gamified version achieved more accuracy on the solutions. In addition, Kao et al. developed a VR programming game where players solve coding challenges by manipulating virtual objects and interacting with a visually stimulating interface \cite{kao2020hack}.

\subsection{Gamified Approaches Towards Electronics Learning}

Similar to programming learning, researchers have used gamified approaches to teach various electronics concepts \cite{luthon2014laborem,salas2022construction,duran2018masterengineer}. Luthon and Larroque developed a remote laboratory called LaboREM that offers a virtual environment where users can engage in practical electronics experiments and simulations through an interactive, gamified interface \cite{luthon2014laborem}. Another study developed a web-based Digital Game for the teaching-learning process on Electronics (DGE) for electronic and electrical engineering students and found that the game positively influenced the students’ assimilation of knowledge and skills development \cite{salas2022construction}. In addition, Duran et al. introduced "MasterEngineer", an educational approach designed for teaching power electronics and drives using a game-based technique that offers an interactive and engaging platform where students can learn complex engineering concepts through a game format \cite{duran2018masterengineer}.

\subsection{Gamified Approaches Integrating Both Programming and Electronics Learning}

In the context of robotics education, it is crucial to incorporate both programming and electronics (two major components of a robotic system) together to provide students with a holistic understanding of these domains. However, research on integrating both programming and electronics within a gamified platform for robotics learning is limited \cite{watanabe2018compulsory,panskyi2021holistic}. For example, Watanabe et al. introduced four different game-based robot contests incorporating elements of electronics and programming to help students in learning how to construct a robot system \cite{watanabe2018compulsory}. Moreover, Panskyi and Rowinska proposed a Digital Game-Based Learning (DGBL) approach that teaches programming, electronics, and robotics through three educational phases \cite{panskyi2021holistic}. The initial programming teaching phase uses the Scratch visual programming environment \cite{resnick2009scratch} while the electronics and robotics teaching phases utilize hands-on training through mobile robots and microcontrollers. Nevertheless, these approaches focused more on the teaching aspect rather than the enjoyment component and may be difficult to implement for individuals with little to no experience in these fields. As a result, I designed an easily accessible, easy-to-play, and enjoyable game with great replayability that allows players to get familiar with robotics, electronics, and programming.

\begin{figure*}[h]
  \centering
  \includegraphics[width=\linewidth]{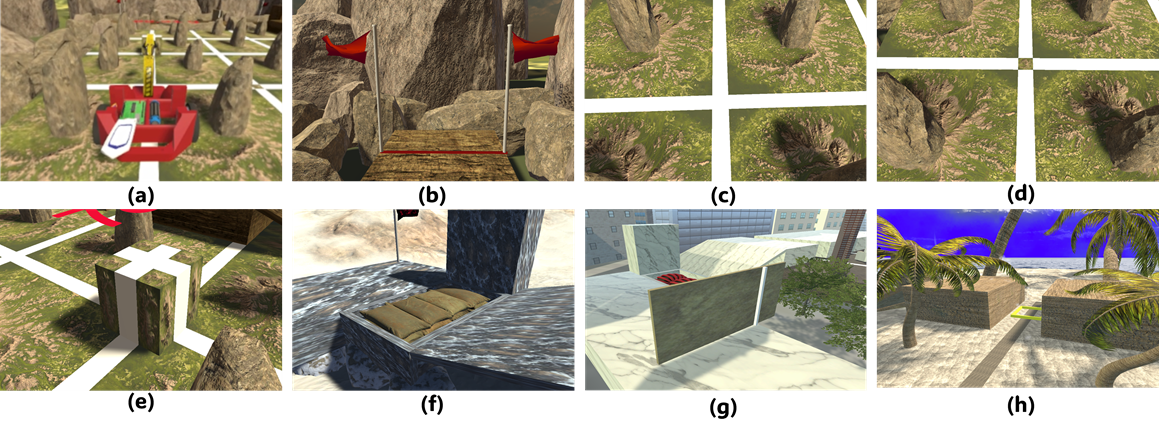}
  \caption{Examples of the main gameplay elements, including (a) starting point, (b) ending point, (c) "legal" nodes, (d) "illegal" nodes, and (e) pickup object. Some miscellaneous elements are also shown, such as (f) sandbag track, (g) rotating gate, and (h) bridge with deposit point}
  \Description{Screenshots showing the the starting point, ending point, legal nodes, illegal nodes, pickup object. Some other gameplay elements are also shown, such as sandbag track, rotating gate and bridge.}
  \label{fig:game_01}
\end{figure*}

\begin{figure*}[h]
  \centering
  \includegraphics[width=\linewidth]{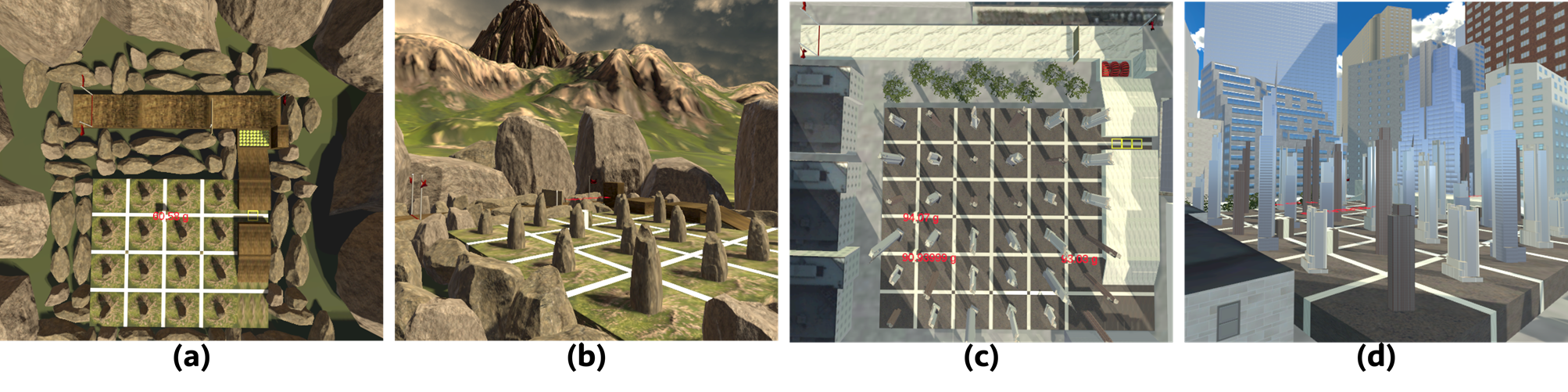}
  \caption{Two versions of gameplay maps of different difficulty containing “illegal” nodes, pickup objects, and other obstacles in different positions, while providing at least one accessible path, for: (a, b) easy difficulty "South American Mountains" map with top and side views, and (c, d) hard difficulty "New York" map with top and side views}
  \Description{Two examples of gameplay maps, "South American Mountains" and "New York". Screenshots of both top and side views are included.}
  \label{fig:levels_01}
\end{figure*}

\section{The Game}

\subsection{Gameplay}

\subsubsection{Overview}

\textit{GeoBotsVR} is set in a sci-fi setting in the year 3000, where players control two-wheeled robots (inspired by early 1980-90s robots) navigating various locations on Earth (Fig. \ref{fig:levels_01} and Fig. \ref{fig:arcade_03}a and b). The grid-based playable area (Fig. \ref{fig:levels_01}) involves "legal" (Fig. \ref{fig:game_01}c) and "illegal" nodes (Fig. \ref{fig:game_01}d), with tasks such as picking up objects and overcoming obstacles (Fig. \ref{fig:game_01}e). The game also includes additional elements such as sandbag tracks and bridges (Fig. \ref{fig:game_01}f, g, and h), offering diverse challenges. Players must drive their robot from the start point (Fig. \ref{fig:game_01}a), follow legal nodes, avoid illegal ones, complete various objectives, and progress to the endpoint (Fig. \ref{fig:game_01}b). Illegal maneuvers (such as hitting an “illegal node”) prompt immediate restarts. The main goal is to complete levels efficiently, earning in-game currency based on time to completion and nodes navigated. The game also contains a tutorial explaining different gameplay elements for first-time players.

\subsubsection{Maps}

The game features nine different maps (based on various locations around the world) with three difficulty levels (easy, medium, and hard) (Fig. \ref{fig:levels_01}). Procedural Content Generation (PCG) using Dijkstra's algorithm generates grid structures with "legal" and "illegal" nodes based on the difficulty level (Fig. \ref{fig:levels_01}). The PCG is constrained by a range of the number of "illegal" nodes allowed under the difficulty settings and always ensures a unique level with at least one accessible path. This allows each game level to be unique and provides an endless array of puzzles the player can attempt to solve, enhancing replayability and preventing monotony.

\subsubsection{Presence and locations of gameplay elements}

Based on the difficulty of the chosen map, the game randomly modifies the number and weight of objects to be picked up in the playable area, while ensuring compatibility with the grid map. Moreover, the game uses Dijkstra’s algorithm to ensure that all objects are accessible in the grid map. Similarly, the game randomly generates other gameplay elements (such as those shown in Fig. \ref{fig:game_01}f, g, and h) and randomizes their locations inside the playable area while maintaining compatibility with the grid structure and gameplay objectives. Based on the presence/absence and locations of these gameplay elements and the weight and number of pickup objects, players may have to customize their robot.

\begin{figure*}[h]
  \centering
  \includegraphics[width=\linewidth]{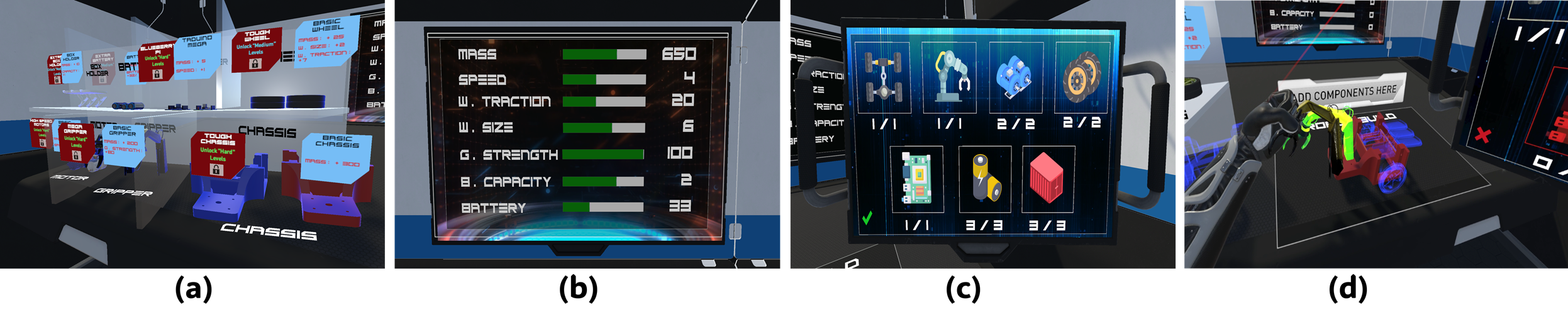}
  \caption{Components of The Workshop, including (a) various robotic components with their information, (b) the robot stats display, and (c) the build progress display with all required components added. An example of adding robotic components to the robot is also shown (d), where an indicator on the robot's body guides the player on where to add the component}
  \Description{Screenshots showing some components of the workshop, such as the robotic components, the robot stats display, and the build progress display. An example of manipulating the robotic components is also shown.}
  \label{fig:workshop_02}
\end{figure*}

\begin{figure*}[h]
  \centering
  \includegraphics[width=\linewidth]{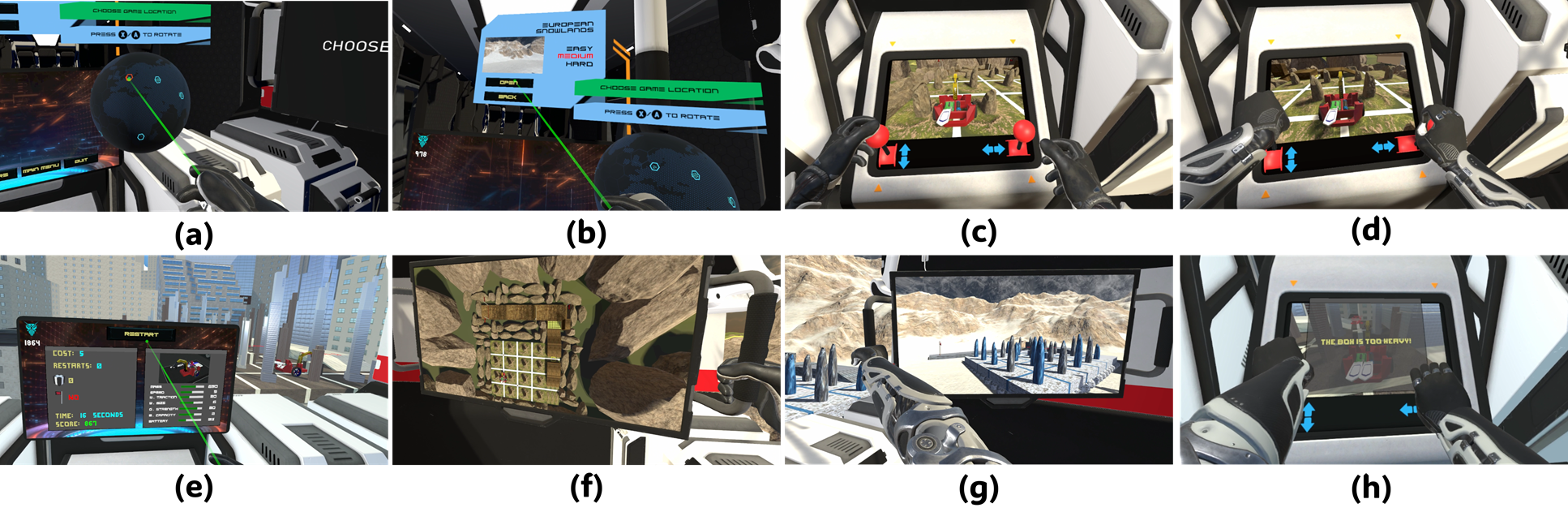}
  \caption{Screenshots of various components of The Arcade, including (a, b) the globe menu for selecting a map, (c, d) joysticks for driving the robot, (e) the dock display showing in-game information, (f) a display showing the top view of the map, (g) a display showing the side view of the map, and (h) example of a warning message}
  \Description{Screenshots showing various components of the arcade, such as the globe menu, joysticks for driving the robot, the dock display, the top and side view displays, and an example of a warning message.}
  \label{fig:arcade_03}
\end{figure*}

\subsection{The Workshop: Robot Component Editor}

\textit{The Workshop} is aimed at familiarizing the player with various components required for the functioning of two-wheeled mobile robots in real life (Fig. \ref{fig:teaser}a). The player’s robot has seven different stats that can affect the gameplay, namely \textit{mass}, \textit{speed}, \textit{wheel traction}, \textit{wheel size}, \textit{grip strength}, \textit{box capacity}, and \textit{battery capacity} (Fig. \ref{fig:workshop_02}b). In \textit{The Workshop}, players can add/remove different components to the robot (Fig. \ref{fig:workshop_02}a), with different combinations of components providing different stats. For example, \textit{motor} and \textit{wheel} components will mostly affect the \textit{speed}, \textit{wheel traction}, and \textit{wheel size} stats of the robot, the \textit{gripper} component will affect the \textit{grip strength} stat, the \textit{box holder} component will affect the \textit{box capacity} stat of the robot, etc. In-game currency can be used to upgrade and/or purchase new components. Additionally, a robot stats display shows the current stats of the robot based on the current components added (Fig. \ref{fig:workshop_02}b). There is also a build progress display that provides information to the players about the type of components currently added to the robot (Fig. \ref{fig:workshop_02}c). Besides showing how to build a robot using different components, \textit{The Workshop} also guides players in adapting the robot to meet specific gameplay objectives. For example, if the playable area contains a sandbag track that needs higher \textit{wheel traction} and \textit{wheel size} to overcome, the player can equip the robot with a better \textit{wheel} component. Similarly, an object in a “medium” difficulty map may be too heavy for the robot to pick up. In that case, the player can attach a better \textit{gripper} component to the robot.

\subsection{The Arcade: Main Gameplay}

Players participate in the main gameplay using various components in \textit{The Arcade} (Fig. \ref{fig:teaser}b), which was designed by blending futuristic elements with 1980-90s arcade elements. Firstly, the player selects the game map from a globe menu simulating the Earth and nine different locations around the world (South American Mountains, African Desert, New York, East Coast Islands, European Snowlands, Russian Gulag, Asian Rainforest, Chinese Temple, and Japanese Industry) (Fig. \ref{fig:arcade_03}a and b). After a map has been selected, the player uses the joysticks on the arcade machine to navigate their robot around the generated environment (Fig. \ref{fig:arcade_03}c and d). There is a dock display which provides in-game information, such as score, time, robot stats, etc. (Fig. \ref{fig:arcade_03}e). Various other components of \textit{The Arcade} assist the player in completing the level, such as the top view display (providing a top-down view of the environment) (Fig. \ref{fig:arcade_03}f), the side view display (providing an angled side view of the environment) (Fig. \ref{fig:arcade_03}g), and various warning messages (Fig. \ref{fig:arcade_03}h).

\begin{figure*}[h]
  \centering
  \includegraphics[width=\linewidth]{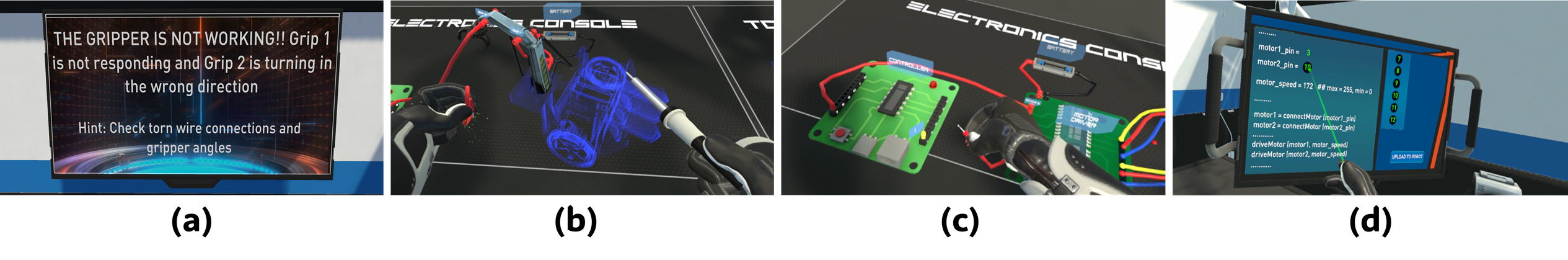}
  \caption{Components and manipulation methods of The Repair Zone: (a) a display showing the problem statement and hints, (b) direct manipulation of electronics using hands with tools and (c) without tools, and (d) manipulating the code using a simple point-and-click method}
  \Description{Screenshots showing some components and manipulation methods of the repair zone. A display showing the problem statement and hints is shown. Moreover, some screenshots show how electronic components are manipulated using tools and without tools. Manipulation of the code using point-and-click is also shown.}
  \label{fig:repair_01}
\end{figure*}

\subsection{The Repair Zone: Programming and Electronics Problems}

Before playing each level, players were sometimes (randomly selected) alerted about some errors with their robot’s electronic components and programming code. Players were required to repair their robot by solving one randomly generated problem (including both electronics and programming components) by going to the \textit{The Repair Zone} (Fig. \ref{fig:teaser}c). The main components of \textit{The Repair Zone} are an electronics console where players can manipulate the electronic components of the robot related to the problem (Fig. \ref{fig:teaser}c), a problem statement display showing hints and information about the problem (Fig. \ref{fig:repair_01}a), and a programming console where players can modify certain sections of their robot’s code (Fig. \ref{fig:teaser}c). Interaction with the electronics and the programming console is also simple and intuitive as players can directly manipulate the electronics with their hands (Fig. \ref{fig:repair_01}b and c) and modify the code using a simple point-and-click method (Fig. \ref{fig:repair_01}d). The generated problems scale with difficulty settings and don't require any prior knowledge of electronics or programming, with the game currently offering 27 such problems across three difficulty levels.

\section{Critical Reflection and Conclusion}

\textit{GeoBotsVR} addresses the gap in integrating robotics, electronics, embedded systems, and programming within a gamified framework for beginners. Existing approaches often prioritized teaching complex concepts over enjoyment and focused on individuals with some experience in related fields \cite{watanabe2018compulsory,panskyi2021holistic}. On the contrary, \textit{GeoBotsVR}, comprising \textit{The Workshop} (Fig. \ref{fig:teaser}a), \textit{The Arcade} (Fig. \ref{fig:teaser}b), and \textit{The Repair Zone} (Fig. \ref{fig:teaser}c), offers an immersive, replayable, practical application-based, easy-to-operate, and enjoyable experience. A crucial feature of \textit{GeoBotsVR} is the ability for players to use the robot they construct, modify, and repair in actual gameplay, enhancing their interest and motivation in learning robotics. This hands-on application of the robot contributes to a more engaging and immersive learning experience. Firstly, the intuitive nature of \textit{The Workshop} and its user interfaces allows players with limited technical knowledge and previous experience to easily familiarize themselves with various components that make up a functioning two-wheeled mobile robot (Fig. \ref{fig:workshop_02}). In addition, using own hands in an immersive VR environment to build the robot simulates a hands-on experience similar to building the robot in real life (Fig. \ref{fig:workshop_02}). Secondly, \textit{The Arcade} provides a sense of embodiment to the players while playing the game. For example, the globe menu provides an intuitive and interesting way for players to choose the game map (Fig. \ref{fig:arcade_03}a and b). In addition, the use of the joysticks in the arcade machine to navigate the robot simulates the experience of driving the robot in real life and increases immersion compared to using button and/or controller input (Fig. \ref{fig:arcade_03}c and d). Lastly, \textit{The Repair Zone} exposes players to some common issues in robot development, with simple and intuitive interactions suitable for beginners (Fig. \ref{fig:repair_01}).

With a target audience of individuals inexperienced in robotics, \textit{GeoBotsVR} provides unique learning opportunities compared to traditional educational modes for teaching robotics, programming, and electronics. It does not require extensive physical space, course-based structure, or instructors and offers simplicity, even compared to existing game-based robotics learning approaches \cite{watanabe2018compulsory,panskyi2021holistic}. For instance, individuals can get familiar with robot building and common problems associated with a robot's programming and electronics through a course-based structure and hands-on learning in a physical space such as a classroom. Players can get the same experience in their leisure time in \textit{GeoBotsVR} in the comfort of their home without any added complexity of investing additional time and effort towards attending a class.

While the VR environment in \textit{GeoBotsVR} enhances immersion and player presence, potential negative side effects like simulator sickness and VR sickness can occur with prolonged exposure. These effects are partially mitigated in \textit{GeoBotsVR}, as all interactions involve using the hands in a static position without any abrupt movements; additional mitigation measures include limiting playtime and taking regular breaks. In addition, although the user interfaces in \textit{GeoBotsVR} were designed to be intuitive and user-friendly, they may be overwhelming in the VR environment for first-time users. In future work, I will focus on improving the ease-of-use of the various interfaces. Nevertheless, \textit{GeoBotsVR} has the potential to be an accessible and enjoyable tool for individuals without prior technical knowledge to familiarize themselves with robotics.

\begin{acks}
The author wishes to thank Dominic Kao for contributing with idea conception, guidance, and funding throughout the project.

This research was supported in part by the National Science Foundation under the award number IIS \#2113991. Any opinions, findings, conclusions, or recommendations expressed in this material are those of the author and do not necessarily reflect the views of the National Science Foundation.
\end{acks}

\bibliographystyle{ACM-Reference-Format}
\bibliography{bibliography}


\begin{thebibliography}{33}


\ifx \showCODEN    \undefined \def \showCODEN     #1{\unskip}     \fi
\ifx \showDOI      \undefined \def \showDOI       #1{#1}\fi
\ifx \showISBNx    \undefined \def \showISBNx     #1{\unskip}     \fi
\ifx \showISBNxiii \undefined \def \showISBNxiii  #1{\unskip}     \fi
\ifx \showISSN     \undefined \def \showISSN      #1{\unskip}     \fi
\ifx \showLCCN     \undefined \def \showLCCN      #1{\unskip}     \fi
\ifx \shownote     \undefined \def \shownote      #1{#1}          \fi
\ifx \showarticletitle \undefined \def \showarticletitle #1{#1}   \fi
\ifx \showURL      \undefined \def \showURL       {\relax}        \fi
\providecommand\bibfield[2]{#2}
\providecommand\bibinfo[2]{#2}
\providecommand\natexlab[1]{#1}
\providecommand\showeprint[2][]{arXiv:#2}

\bibitem[Cheon and Grant(2012)]%
        {cheon2012effects}
\bibfield{author}{\bibinfo{person}{Jongpil Cheon} {and} \bibinfo{person}{Michael~M Grant}.} \bibinfo{year}{2012}\natexlab{}.
\newblock \showarticletitle{The effects of metaphorical interface on germane cognitive load in web-based instruction}.
\newblock \bibinfo{journal}{\emph{Educational Technology Research and Development}}  \bibinfo{volume}{60} (\bibinfo{year}{2012}), \bibinfo{pages}{399--420}.
\newblock


\bibitem[Costa(2023)]%
        {costa2023using}
\bibfield{author}{\bibinfo{person}{Joana~M Costa}.} \bibinfo{year}{2023}\natexlab{}.
\newblock \showarticletitle{Using game concepts to improve programming learning: A multi-level meta-analysis}.
\newblock \bibinfo{journal}{\emph{Computer Applications in Engineering Education}} (\bibinfo{year}{2023}).
\newblock


\bibitem[De~Paula et~al\mbox{.}(2018)]%
        {de2018playing}
\bibfield{author}{\bibinfo{person}{Bruno~Henrique De~Paula}, \bibinfo{person}{Andrew Burn}, \bibinfo{person}{Richard Noss}, {and} \bibinfo{person}{Jos{\'e}~Armando Valente}.} \bibinfo{year}{2018}\natexlab{}.
\newblock \showarticletitle{Playing Beowulf: Bridging computational thinking, arts and literature through game-making}.
\newblock \bibinfo{journal}{\emph{International journal of child-computer interaction}}  \bibinfo{volume}{16} (\bibinfo{year}{2018}), \bibinfo{pages}{39--46}.
\newblock


\bibitem[Dede(2009)]%
        {dede2009immersive}
\bibfield{author}{\bibinfo{person}{Chris Dede}.} \bibinfo{year}{2009}\natexlab{}.
\newblock \showarticletitle{Immersive interfaces for engagement and learning}.
\newblock \bibinfo{journal}{\emph{science}} \bibinfo{volume}{323}, \bibinfo{number}{5910} (\bibinfo{year}{2009}), \bibinfo{pages}{66--69}.
\newblock


\bibitem[Dewantara et~al\mbox{.}(2021)]%
        {dewantara2021game}
\bibfield{author}{\bibinfo{person}{D Dewantara}, \bibinfo{person}{M Misbah}, \bibinfo{person}{S Haryandi}, {and} \bibinfo{person}{S Mahtari}.} \bibinfo{year}{2021}\natexlab{}.
\newblock \showarticletitle{Game-based learning for the mastery of HOTS in prospective physics teachers in digital electronics courses}. In \bibinfo{booktitle}{\emph{Journal of Physics: Conference Series}}, Vol.~\bibinfo{volume}{1869}. IOP Publishing, \bibinfo{pages}{012153}.
\newblock


\bibitem[Duran et~al\mbox{.}(2018)]%
        {duran2018masterengineer}
\bibfield{author}{\bibinfo{person}{MJ Duran}, \bibinfo{person}{I Gonzalez}, \bibinfo{person}{P Garcia-Entrambasaguas}, \bibinfo{person}{JJ Aciego}, \bibinfo{person}{A Gonzalez}, {and} \bibinfo{person}{N Rios}.} \bibinfo{year}{2018}\natexlab{}.
\newblock \showarticletitle{MasterEngineer: A game-based technique in power electronics and drives teaching}. In \bibinfo{booktitle}{\emph{2018 XIII Technologies Applied to Electronics Teaching Conference (TAEE)}}. IEEE, \bibinfo{pages}{1--6}.
\newblock


\bibitem[Eguchi(2014)]%
        {eguchi2014robotics}
\bibfield{author}{\bibinfo{person}{Amy Eguchi}.} \bibinfo{year}{2014}\natexlab{}.
\newblock \showarticletitle{Robotics as a learning tool for educational transformation}. In \bibinfo{booktitle}{\emph{Proceedings of 4th international workshop teaching robotics, teaching with robotics \& 5th international conference robotics in education}}, Vol.~\bibinfo{volume}{18}. \bibinfo{pages}{27--34}.
\newblock


\bibitem[Hussein and N{\"a}tterdal(2015)]%
        {hussein2015benefits}
\bibfield{author}{\bibinfo{person}{Mustafa Hussein} {and} \bibinfo{person}{Carl N{\"a}tterdal}.} \bibinfo{year}{2015}\natexlab{}.
\newblock \showarticletitle{The benefits of virtual reality in education-A comparision Study}.
\newblock  (\bibinfo{year}{2015}).
\newblock


\bibitem[Kao et~al\mbox{.}(2020)]%
        {kao2020hack}
\bibfield{author}{\bibinfo{person}{Dominic Kao}, \bibinfo{person}{Christos Mousas}, \bibinfo{person}{Alejandra~J Magana}, \bibinfo{person}{D~Fox Harrell}, \bibinfo{person}{Rabindra Ratan}, \bibinfo{person}{Edward~F Melcer}, \bibinfo{person}{Brett Sherrick}, \bibinfo{person}{Paul Parsons}, {and} \bibinfo{person}{Dmitri~A Gusev}.} \bibinfo{year}{2020}\natexlab{}.
\newblock \showarticletitle{Hack. VR: A programming game in virtual reality}.
\newblock \bibinfo{journal}{\emph{arXiv preprint arXiv:2007.04495}} (\bibinfo{year}{2020}).
\newblock


\bibitem[Lee and Wong(2014)]%
        {lee2014learning}
\bibfield{author}{\bibinfo{person}{Elinda Ai-Lim Lee} {and} \bibinfo{person}{Kok~Wai Wong}.} \bibinfo{year}{2014}\natexlab{}.
\newblock \showarticletitle{Learning with desktop virtual reality: Low spatial ability learners are more positively affected}.
\newblock \bibinfo{journal}{\emph{Computers \& Education}}  \bibinfo{volume}{79} (\bibinfo{year}{2014}), \bibinfo{pages}{49--58}.
\newblock


\bibitem[Lee et~al\mbox{.}(2014)]%
        {lee2014principles}
\bibfield{author}{\bibinfo{person}{Michael~J Lee}, \bibinfo{person}{Faezeh Bahmani}, \bibinfo{person}{Irwin Kwan}, \bibinfo{person}{Jilian LaFerte}, \bibinfo{person}{Polina Charters}, \bibinfo{person}{Amber Horvath}, \bibinfo{person}{Fanny Luor}, \bibinfo{person}{Jill Cao}, \bibinfo{person}{Catherine Law}, \bibinfo{person}{Michael Beswetherick}, {et~al\mbox{.}}} \bibinfo{year}{2014}\natexlab{}.
\newblock \showarticletitle{Principles of a debugging-first puzzle game for computing education}. In \bibinfo{booktitle}{\emph{2014 IEEE symposium on visual languages and human-centric computing (VL/HCC)}}. IEEE, \bibinfo{pages}{57--64}.
\newblock


\bibitem[Lee and Ko(2015)]%
        {lee2015comparing}
\bibfield{author}{\bibinfo{person}{Michael~J Lee} {and} \bibinfo{person}{Amy~J Ko}.} \bibinfo{year}{2015}\natexlab{}.
\newblock \showarticletitle{Comparing the effectiveness of online learning approaches on CS1 learning outcomes}. In \bibinfo{booktitle}{\emph{Proceedings of the eleventh annual international conference on international computing education research}}. \bibinfo{pages}{237--246}.
\newblock


\bibitem[Linnenbrink and Pintrich(2002)]%
        {linnenbrink2002motivation}
\bibfield{author}{\bibinfo{person}{Elizabeth~A Linnenbrink} {and} \bibinfo{person}{Paul~R Pintrich}.} \bibinfo{year}{2002}\natexlab{}.
\newblock \showarticletitle{Motivation as an enabler for academic success}.
\newblock \bibinfo{journal}{\emph{School psychology review}} \bibinfo{volume}{31}, \bibinfo{number}{3} (\bibinfo{year}{2002}), \bibinfo{pages}{313--327}.
\newblock


\bibitem[Luthon and Larroque(2014)]%
        {luthon2014laborem}
\bibfield{author}{\bibinfo{person}{Franck Luthon} {and} \bibinfo{person}{Benoit Larroque}.} \bibinfo{year}{2014}\natexlab{}.
\newblock \showarticletitle{LaboREM—A remote laboratory for game-like training in electronics}.
\newblock \bibinfo{journal}{\emph{IEEE Transactions on learning technologies}} \bibinfo{volume}{8}, \bibinfo{number}{3} (\bibinfo{year}{2014}), \bibinfo{pages}{311--321}.
\newblock


\bibitem[Lynch(2006)]%
        {lynch2006motivational}
\bibfield{author}{\bibinfo{person}{Douglas~J Lynch}.} \bibinfo{year}{2006}\natexlab{}.
\newblock \showarticletitle{Motivational factors, learning strategies and resource management as predictors of course grades}.
\newblock \bibinfo{journal}{\emph{College Student Journal}} \bibinfo{volume}{40}, \bibinfo{number}{2} (\bibinfo{year}{2006}), \bibinfo{pages}{423--429}.
\newblock


\bibitem[Melcer(2018)]%
        {melcer2018learning}
\bibfield{author}{\bibinfo{person}{Edward~F Melcer}.} \bibinfo{year}{2018}\natexlab{}.
\newblock \emph{\bibinfo{title}{Learning with the body: understanding the design space of embodied educational technology}}.
\newblock \bibinfo{thesistype}{Ph.\,D. Dissertation}. \bibinfo{school}{New York University Tandon School of Engineering}.
\newblock


\bibitem[Mouton and Blake(1984)]%
        {mouton1984synergogy}
\bibfield{author}{\bibinfo{person}{Jane~Srygley Mouton} {and} \bibinfo{person}{Robert~R Blake}.} \bibinfo{year}{1984}\natexlab{}.
\newblock \bibinfo{booktitle}{\emph{Synergogy: A New Strategy for Education, Training, and Development.}}
\newblock \bibinfo{publisher}{ERIC}.
\newblock


\bibitem[Mubarrat et~al\mbox{.}(2022a)]%
        {Chowdhury_2022_physics}
\bibfield{author}{\bibinfo{person}{Syed~T. Mubarrat}, \bibinfo{person}{Suman Chowdhury}, {and} \bibinfo{person}{Antonio Fernandes}.} \bibinfo{year}{2022}\natexlab{a}.
\newblock \showarticletitle{A Physics-based Virtual Reality System Design and Evaluation by Simulating Human-Robot Collaboration}.
\newblock  (\bibinfo{date}{Jan.} \bibinfo{year}{2022}).
\newblock
\urldef\tempurl%
\url{https://doi.org/10.36227/techrxiv.18972773.v1}
\showDOI{\tempurl}


\bibitem[Mubarrat et~al\mbox{.}(2022b)]%
        {Chowdhury_2022_coreg}
\bibfield{author}{\bibinfo{person}{Syed~T. Mubarrat}, \bibinfo{person}{Suman Chowdhury}, \bibinfo{person}{Antonio Fernandes}, {and} \bibinfo{person}{Kieran Binkley}.} \bibinfo{year}{2022}\natexlab{b}.
\newblock \showarticletitle{Evaluating Visual-Spatiotemporal Coregistration of a Physics-based Virtual Reality Haptic Interface}.
\newblock  (\bibinfo{date}{March} \bibinfo{year}{2022}).
\newblock
\urldef\tempurl%
\url{https://doi.org/10.36227/techrxiv.19242951.v1}
\showDOI{\tempurl}


\bibitem[Mubarrat et~al\mbox{.}(2020)]%
        {mubarrat2020evaluation}
\bibfield{author}{\bibinfo{person}{Syed~T Mubarrat}, \bibinfo{person}{Oluwatosin Opafunso}, {and} \bibinfo{person}{Suman~K Chowdhury}.} \bibinfo{year}{2020}\natexlab{}.
\newblock \showarticletitle{The Evaluation of User Experience and Functional Workload of a Physically Inter-active Virtual Reality System}. In \bibinfo{booktitle}{\emph{Proceedings of the Human Factors and Ergonomics Society Annual Meeting}}, Vol.~\bibinfo{volume}{64}. SAGE Publications Sage CA: Los Angeles, CA, \bibinfo{pages}{2084--2086}.
\newblock


\bibitem[Panskyi and ROWI{\'N}SKA(2021)]%
        {panskyi2021holistic}
\bibfield{author}{\bibinfo{person}{Taras Panskyi} {and} \bibinfo{person}{Zdzis{\l}awa ROWI{\'N}SKA}.} \bibinfo{year}{2021}\natexlab{}.
\newblock \showarticletitle{A Holistic Digital Game-Based Learning Approach to Out-of-School Primary Programming Education.}
\newblock \bibinfo{journal}{\emph{Informatics in Education}} \bibinfo{volume}{20}, \bibinfo{number}{2} (\bibinfo{year}{2021}).
\newblock


\bibitem[Ratan(2013)]%
        {ratan2013self}
\bibfield{author}{\bibinfo{person}{Rabindra Ratan}.} \bibinfo{year}{2013}\natexlab{}.
\newblock \showarticletitle{Self-presence, explicated: Body, emotion, and identity extension into the virtual self}.
\newblock In \bibinfo{booktitle}{\emph{Handbook of research on technoself: Identity in a technological society}}. \bibinfo{publisher}{IGI Global}, \bibinfo{pages}{322--336}.
\newblock


\bibitem[Resnick et~al\mbox{.}(2009)]%
        {resnick2009scratch}
\bibfield{author}{\bibinfo{person}{Mitchel Resnick}, \bibinfo{person}{John Maloney}, \bibinfo{person}{Andr{\'e}s Monroy-Hern{\'a}ndez}, \bibinfo{person}{Natalie Rusk}, \bibinfo{person}{Evelyn Eastmond}, \bibinfo{person}{Karen Brennan}, \bibinfo{person}{Amon Millner}, \bibinfo{person}{Eric Rosenbaum}, \bibinfo{person}{Jay Silver}, \bibinfo{person}{Brian Silverman}, {et~al\mbox{.}}} \bibinfo{year}{2009}\natexlab{}.
\newblock \showarticletitle{Scratch: programming for all}.
\newblock \bibinfo{journal}{\emph{Commun. ACM}} \bibinfo{volume}{52}, \bibinfo{number}{11} (\bibinfo{year}{2009}), \bibinfo{pages}{60--67}.
\newblock


\bibitem[Rockland et~al\mbox{.}(2010)]%
        {rockland_advancing_2010}
\bibfield{author}{\bibinfo{person}{Ronald Rockland}, \bibinfo{person}{Diane~S Bloom}, \bibinfo{person}{John Carpinelli}, \bibinfo{person}{Levelle Burr-Alexander}, \bibinfo{person}{Linda~S Hirsch}, {and} \bibinfo{person}{Howard Kimmel}.} \bibinfo{year}{2010}\natexlab{}.
\newblock \showarticletitle{Advancing the "{E}" in {K}-12 {STEM} {Education}}.
\newblock \bibinfo{journal}{\emph{Journal of Technology Studies}} \bibinfo{volume}{36}, \bibinfo{number}{1} (\bibinfo{year}{2010}), \bibinfo{pages}{53--64}.
\newblock
\showISSN{10716084}
\urldef\tempurl%
\url{https://www.proquest.com/scholarly-journals/advancing-e-k-12-stem-education/docview/819727012/se-2?accountid=13360}
\showURL{%
\tempurl}
\newblock
\shownote{Place: Bowling Green Publisher: Epsilon Pi Tau}.


\bibitem[Ryan and Deci(2000)]%
        {ryan2000self}
\bibfield{author}{\bibinfo{person}{Richard~M Ryan} {and} \bibinfo{person}{Edward~L Deci}.} \bibinfo{year}{2000}\natexlab{}.
\newblock \showarticletitle{Self-determination theory and the facilitation of intrinsic motivation, social development, and well-being.}
\newblock \bibinfo{journal}{\emph{American psychologist}} \bibinfo{volume}{55}, \bibinfo{number}{1} (\bibinfo{year}{2000}), \bibinfo{pages}{68}.
\newblock


\bibitem[Salas-Rueda et~al\mbox{.}(2022)]%
        {salas2022construction}
\bibfield{author}{\bibinfo{person}{Ricardo-Ad{\'a}n Salas-Rueda}, \bibinfo{person}{Clara Alvarado-Zamorano}, {and} \bibinfo{person}{Jes{\'u}s Ram{\'\i}rez-Ortega}.} \bibinfo{year}{2022}\natexlab{}.
\newblock \showarticletitle{Construction of a Web Game for the Teaching-Learning Process of Electronics during the COVID-19 Pandemic.}
\newblock \bibinfo{journal}{\emph{Educational Process: International Journal}} \bibinfo{volume}{11}, \bibinfo{number}{2} (\bibinfo{year}{2022}), \bibinfo{pages}{130--146}.
\newblock


\bibitem[Shafie and Abdullah(2020)]%
        {shafie2020gamification}
\bibfield{author}{\bibinfo{person}{Azlinda Shafie} {and} \bibinfo{person}{Zaleha Abdullah}.} \bibinfo{year}{2020}\natexlab{}.
\newblock \showarticletitle{Gamification in learning programming language}. In \bibinfo{booktitle}{\emph{SA Conference Series: Industrial Revolution 4.0}}, Vol.~\bibinfo{volume}{1}. \bibinfo{pages}{181--187}.
\newblock


\bibitem[Shin(2017)]%
        {shin2017role}
\bibfield{author}{\bibinfo{person}{Dong-Hee Shin}.} \bibinfo{year}{2017}\natexlab{}.
\newblock \showarticletitle{The role of affordance in the experience of virtual reality learning: Technological and affective affordances in virtual reality}.
\newblock \bibinfo{journal}{\emph{Telematics and Informatics}} \bibinfo{volume}{34}, \bibinfo{number}{8} (\bibinfo{year}{2017}), \bibinfo{pages}{1826--1836}.
\newblock


\bibitem[Smiderle et~al\mbox{.}(2020)]%
        {smiderle2020impact}
\bibfield{author}{\bibinfo{person}{Rodrigo Smiderle}, \bibinfo{person}{Sandro~Jos{\'e} Rigo}, \bibinfo{person}{Leonardo~B Marques}, \bibinfo{person}{Jorge~Arthur Pe{\c{c}}anha~de Miranda~Coelho}, {and} \bibinfo{person}{Patricia~A Jaques}.} \bibinfo{year}{2020}\natexlab{}.
\newblock \showarticletitle{The impact of gamification on students’ learning, engagement and behavior based on their personality traits}.
\newblock \bibinfo{journal}{\emph{Smart Learning Environments}} \bibinfo{volume}{7}, \bibinfo{number}{1} (\bibinfo{year}{2020}), \bibinfo{pages}{1--11}.
\newblock


\bibitem[Srinivasan et~al\mbox{.}(2021)]%
        {srinivasan2021biomechanical}
\bibfield{author}{\bibinfo{person}{Manoj Srinivasan}, \bibinfo{person}{Syed~T Mubarrat}, \bibinfo{person}{Quentin Humphrey}, \bibinfo{person}{Thomas Chen}, \bibinfo{person}{Kieran Binkley}, {and} \bibinfo{person}{Suman~K Chowdhury}.} \bibinfo{year}{2021}\natexlab{}.
\newblock \showarticletitle{The Biomechanical Evaluation of a Human-Robot Collaborative Task in a Physically Interactive Virtual Reality Simulation Testbed}. In \bibinfo{booktitle}{\emph{Proceedings of the Human Factors and Ergonomics Society Annual Meeting}}, Vol.~\bibinfo{volume}{65}. SAGE Publications Sage CA: Los Angeles, CA, \bibinfo{pages}{403--407}.
\newblock


\bibitem[Steed et~al\mbox{.}(2016)]%
        {steed2016impact}
\bibfield{author}{\bibinfo{person}{Anthony Steed}, \bibinfo{person}{Ye Pan}, \bibinfo{person}{Fiona Zisch}, {and} \bibinfo{person}{William Steptoe}.} \bibinfo{year}{2016}\natexlab{}.
\newblock \showarticletitle{The impact of a self-avatar on cognitive load in immersive virtual reality}. In \bibinfo{booktitle}{\emph{2016 IEEE virtual reality (VR)}}. IEEE, \bibinfo{pages}{67--76}.
\newblock


\bibitem[Watanabe et~al\mbox{.}(2018)]%
        {watanabe2018compulsory}
\bibfield{author}{\bibinfo{person}{Harumi Watanabe}, \bibinfo{person}{Mikiko Sato}, \bibinfo{person}{Masafumi Miwa}, \bibinfo{person}{Makoto Imamura}, \bibinfo{person}{Shintaro Hosoai}, \bibinfo{person}{Nobuhiko Ogura}, \bibinfo{person}{Hiroyuki Nakamura}, {and} \bibinfo{person}{Kenji Hisazumi}.} \bibinfo{year}{2018}\natexlab{}.
\newblock \showarticletitle{Compulsory game based robot contest for embedded system development education}. In \bibinfo{booktitle}{\emph{Proceedings of the 2018 7th International Conference on Software and Computer Applications}}. \bibinfo{pages}{259--263}.
\newblock


\bibitem[Zeidler(2016)]%
        {zeidler2016stem}
\bibfield{author}{\bibinfo{person}{Dana~L Zeidler}.} \bibinfo{year}{2016}\natexlab{}.
\newblock \showarticletitle{STEM education: A deficit framework for the twenty first century? A sociocultural socioscientific response}.
\newblock \bibinfo{journal}{\emph{Cultural Studies of Science Education}}  \bibinfo{volume}{11} (\bibinfo{year}{2016}), \bibinfo{pages}{11--26}.
\newblock


\end{thebibliography}

\end{document}